\documentclass[prd,twocolumn,amsmath,amssymb,floatfix,nofootinbib]{revtex4}
\usepackage{mathrsfs,bm}
\usepackage{longtable,lscape}
\usepackage{txfonts}
\usepackage{indentfirst}
\usepackage{graphicx,color,dcolumn,booktabs}
\usepackage{multirow}
\usepackage{indentfirst}
\usepackage{multirow}
\usepackage{epsfig}
\usepackage{graphicx,color,dcolumn}
\usepackage{epstopdf}
\usepackage{amsmath}
\usepackage{diagbox}
\usepackage{changes}
\usepackage{threeparttable}
\usepackage{booktabs}
\usepackage{color}
\usepackage{amssymb}
\definecolor{cover}{rgb}{0.77,0.87,0.88}
\definecolor{blueone}{rgb}{0.1,0.1,.7}
\definecolor{citec}{rgb}{0.14,0.47,0.09}
\definecolor{two}{rgb}{0.0,0.5,0.}
\definecolor{three}{rgb}{.5,.1,0.15}
\usepackage[bookmarks=true,bookmarksopen=false,plainpages=false,breaklinks=true,
   bookmarksnumbered=true,hypertexnames=false,
   filecolor=blue,urlcolor=three,menucolor=three,
   linkcolor=three,citecolor=blueone, colorlinks,
   anchorcolor=blue,runcolor=pink,frenchlinks=red
   pdfstartview=FitH,pdftitle=title,%
   pdfauthor=author]{hyperref}

\usepackage{relsize}
\usepackage{xspace}
\def\babar{\mbox{\slshape B\kern-0.1em{\smaller A}\kern-0.1em
    B\kern-0.1em{\smaller A\kern-0.2em R}}}

\newcolumntype{C}{>{$}c<{$}}
\AtBeginDocument{
\heavyrulewidth=.08em
\lightrulewidth=.05em
\cmidrulewidth=.03em
\belowrulesep=.65ex
\belowbottomsep=0pt
\aboverulesep=.4ex
\abovetopsep=0pt
\cmidrulesep=\doublerulesep
\cmidrulekern=.5em
\defaultaddspace=.5em
}

\begin{document}
\title{Heavy-strange meson molecules and possible candidates $D^*_{s0}(2317)$, $D_{s1}(2460)$, and $X_0(2900)$}
\author{Shu-Yi Kong$^1$, Jun-Tao Zhu$^1$, Dan Song$^1$, Jun He$^{1,2}$\footnote{Corresponding author: junhe@njnu.edu.cn}}

\affiliation{$^1$Department of  Physics and Institute of Theoretical Physics, Nanjing Normal University, Nanjing 210097, China\\
$^2$Lanzhou Center for Theoretical Physics, Lanzhou University, Lanzhou 730000, China}

\date{\today}
\begin{abstract}
In this work, we systematically investigate the heavy-strange meson systems, $D^{(*)}K^{(*)}/\bar{B}^{(*)}K^{(*)}$ and $\bar{D}^{(*)}K^{(*)}/B^{(*)}K^{(*)}$, to study  possible molecules in a quasipotenial Bethe-Salpter equation approach together with the one-boson exchange model. The potential is achieved with the help of the hidden-gauge Lagrangians.
 Molecular states are found from all six $S$-wave isoscalar interactions of $D^{(*)}K^{(*)}$ or $\bar{B}^{(*)}K^{(*)}$.  The charmed-strange mesons $D^*_{s0}(2317)$ and $D_{s1}(2460)$ can be related to the ${D}K$ and $D^*K$ states with spin parities $0^+$ and $1^+$, respectively. In the current model, the $\bar{B}K^*$ molecular state with $1^+$ is the best candidate of the recent observed $B_{sJ}(6158)$.  Four molecular states are produced from the interactions of  $\bar{D}^{(*)}K^{(*)}$ or $B^{(*)}K^{(*)}$. The relation between the $\bar{D}^*{K}^*$  molecular state with $0^+$ and  the $X_0(2900)$ is also discussed. No isovector molecular states are found from the interactions considered. The current results are helpful to understand the internal structure of $D^*_{s0}(2317)$, $D_{s1}(2460)$, $X_0(2900)$, and new $B_{sJ}$ states. The experimental research for more heavy-strange meson molecules  is suggested.
\end{abstract}

\maketitle

\section{INTRODUCTION}

Recently, the LHCb collaboration reported a new narrow peak $X_0(2900)$ with a mass of $2866\pm7$~MeV and a width of $57.2 \pm 12.9$~MeV in the $D^-K^+$ invariant mass distribution~\cite{Aaij:2020hon,Aaij:2020ypa}.  Considering its decay channel, it should  contain four different flavor quarks $\bar{c}ud\bar{s}$, and therefore draws the attention of the community. Compact tetraquark picture was proposed to interpret this state ~\cite{Zhang:2020oze,Wang:2020xyc,He:2020jna,Wang:2020prk}. However, some studies  disfavor such assignment, such as a study in extended relativized quark model~\cite{Lu:2020qmp}.
Since $X_0(2900)$ is close to the $\bar{D}^*K^*$ threshold, some authors tended to choose the molecule interpretation~\cite{Xue:2020vtq,Liu:2020nil,Chen:2020aos,Agaev:2020nrc,Huang:2020ptc,Mutuk:2020igv,Molina:2020hde,Xiao:2020ltm,He:2020btl}.
Within the molecule scheme, the investigations of $X_0(2900)$ were performed within chiral unitary approach~\cite{Molina:2020hde}, a quark delocalization color-screening model~\cite{Xue:2020vtq}, a one-boson exchange model~\cite{Liu:2020nil}, the QCD sum rule~\cite{Chen:2020aos,Agaev:2020nrc,Mutuk:2020igv}, and an
 effective Lagrangian approach~\cite{Xiao:2020ltm}. In addition to resonant state interpretations, kinematic effects like triangle singularity also might cause the exotic peak found by LHCb Collaboration~\cite{Liu:2020orv,Burns:2020epm}.

In fact, the states near the $D^{(*)}K^{(*)}$ threshold have attracted much attention. As early as 2003, the BaBar collaboration reported a narrow peak $D^*_{s0}(2317)$ near the $DK$ threshold~\cite{Aubert:2003fg}, and later confirmed at CLEO~\cite{Besson:2003cp} and BELLE~\cite{Krokovny:2003zq}. In addition to the $D^*_{s0}(2317)$, the CLEO collaboration  observed another narrow peak $D_{s1}(2460)$ near the $D^*K$ threshold~\cite{Besson:2003cp}.
There is no doubt that theorists would first assign  them as traditional $c\bar{s}$ meson structures as the missing $^{3}P_0$ and $^{1}P_1$ $c\bar{s}$ states, respectively~\cite{Godfrey:2003kg,Rosner:2006jz}. However, this explanation is implausible for two reasons~\cite{Barnes:2003dj}. For one thing, the experimental masses are much higher than previous conventional quark model prediction by Godfrey and Isgur~\cite{Godfrey:1985xj}. For another, considering the heavy quark symmetry, $^{3}P_0$ state is predicted to have width of an order of hundreds of MeV~\cite{Godfrey:1986wj}, which conflicts with the experimental value, $<3.8$~MeV.
Since these  states are not supposed to be the candidates of $c\bar{s}$ mesons family, theoretical physicist investigate the possibility of  exotic explanations naturally. Besides the tetraquark interpretation~\cite{Cheng:2003kg,Chen:2004dy,Kim:2005gt,Nielsen:2005ia,Terasaki:2005kc,Wang:2006uba},  it is popular  to assign these two charm-strange mesons as $DK$ and $D^*K$ molecules, respectively, due to their closeness to the thresholds~\cite{Barnes:2003dj,Navarra:2015iea,Kolomeitsev:2003ac,Hofmann:2003je,
Guo:2006fu,Zhang:2006ix,Rosner:2006vc,Guo:2006rp}.

Up to now, the possible molecular states near the $B^{(*)}K^{(*)}$ threshold are discussed scarcely in the literature. What is worth mentioning is  the open flavor state $X(5568)$ reported by the D0 collaboration in 2016~\cite{D0:2016mwd}. Unfortunately, such  state is too far below the $BK$ threshold to be taken as a molecule. Moreover, it was not confirmed by  other measurements at LHCb, CMS, CDF, and ATLAS~\cite{Aaij:2016iev,Sirunyan:2017ofq,Aaltonen:2017voc,Aaboud:2018hgx}.
Recently, two structures are observed at LHCb, that is, $B_{sJ}(6064)$ and $B_{sJ}(6114)$ if decays directly to the $B^+K^-$ final state, or $B_{sJ}(6109)$ and $B_{sJ}(6158)$ if instead proceeds through $B^{*+}K^-$~\cite{Aaij:2020hcw}. The $B_{sJ}(6158)$ is quite close to the $\bar{B}K^*$ threshold. It is interesting to consider it as a $\bar{B}K^*$ molecular state.

\begin{figure}[h!]
\centering
\includegraphics[scale=0.83,bb=50 100 350 250,clip]{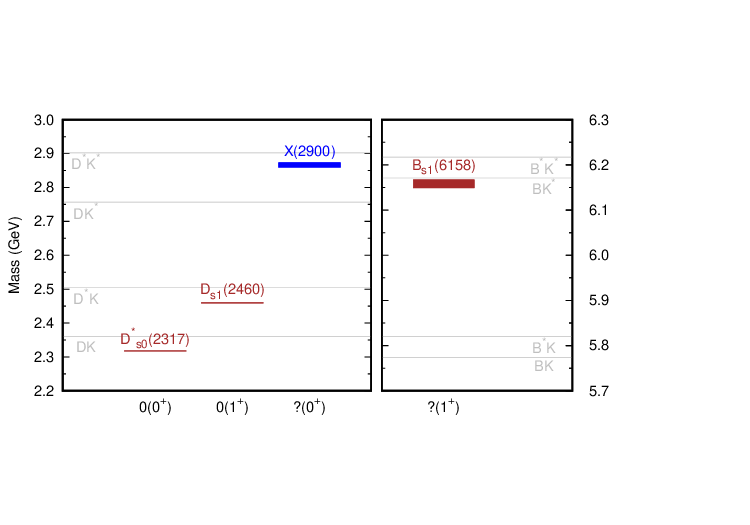}
\caption{Thresholds of $D^{(*)}K^{(*)}/\bar{B}^{(*)}K^{(*)}$ compared with the experimental masses of  $D^*_{s0}(2317)$,~$D_{s1}(2460)$, $X_0(2900)$, and new $B_{sJ}(6158)$ states.}
\label{exp}
\end{figure}

We illustrate all these states with the thresholds of $D^{(*)}K^{(*)}/\bar{B}^{(*)}K^{(*)}$ in Fig.~\ref{exp}. 
One can find that  these states locate at the $DK$, $D^*K$, and $D^*K^*$ thresholds in charmed sector and the $BK^*$ threshold in the bottom sector, respectively. As presented above, the explicit discussions about each experimentally observed states have been performed in the literature. To further understand those states observed in experiment, it is helpful to perform a study of the  interactions  of $D^{(*)}K^{(*)}/\bar{B}^{(*)}K^{(*)}$ and $\bar{D}^{(*)}K^{(*)}/B^{(*)}K^{(*)}$ systematically.

In the literature, there exists some early studies about the possible molecular states composed of (anti-)charmed and strange mesons~\cite{Guo:2011dd,Molina:2010tx}. In our previous work~\cite{He:2020btl}, we studied the $\bar{D}^*K^*$ interaction in a quasipotential Bethe-Salpter equation (qBSE) approach and construct one-boson exchange potential with the help of heavy quark and chiral symmetries. The calculation suggested that $X_0(2900)$ can be explained as a $\bar{D}^*K^*$ molecular state with $J^P=0^+$. In this work, we extend it to do a systemical study on the interaction of charm-strange and bottom-strange meson systems. The bound states are searched for as the pole of the scattering amplitude obtained by solving qBSE. Different from our previous works~\cite{He:2019rva,He:2019csk,He:2014nya,He:2014mja,Ding:2020dio,Zhu:2021lhd,He:2019ify,He:2015cea}, we adopt the  hidden-gauge Lagrangians to construct the potential kernels to make the theoretical frames  more self-consistent than  previous study~\cite{He:2020btl}. All possible molecular states from $S$-wave interactions of $D^{(*)}K^{(*)}/\bar{B}^{(*)}K^{(*)}$ and $\bar{D}^{(*)}K^{(*)}/B^{(*)}K^{(*)}$  will be considered in the calculation. The possible relations between these molecular states and those states observed in experiment will be discussed based on the results.  And, more states will be predicted for further experimental studies.

The paper is organized as follows. After the Introduction, the potential kernels of heavy-strange meson systems are presented, which are obtained with the help of the hidden-gauge Lagrangians, and the qBSE approach is introduced briefly. In Section \ref{Sec: results}, the numerical results of the molecular states from the systems  $D^{(*)}K^{(*)}$ and $\bar{B}^{(*)}K^{(*)}$   with quark configuration $[Q\bar{q}][q\bar{s}]$ and  the systems $\bar{D}^{(*)}K^{(*)}$ and ${B}^{(*)}K^{(*)}$   with quark configurations $[\bar{Q}q][q\bar{s}]$ are presented. The bound states are compared with the experimental results and their relations are discussed. More possible molecular states with other quantum numbers are also predicted.
  In Section \ref{Sec: summary}, discussion and summary are given.

\section{Theoretical frame}\label{Sec: Formalism}

To study the systems considered in the current work, we need to construct the potential of corresponding interactions.  The one-boson exchange model will be adopted with the light mesons $\pi$, $\eta$, $\eta'$, $\rho$, and $\omega$ mediating the interaction between two constituent mesons. To achieve  explicit forms of the interaction potential, we need to describe  vertices of between constituent mesons and the exchanged mesons.  In the following, we introduce the hidden-gauge Lagrangians to perform  construction of  potential.
For the systems considered in the current work, the couplings of  exchanged  light mesons to the charmed/bottom meson and strange meson are required.  Such couplings can be described by the hidden-gauge Lagrangians as ~\cite{Bando:1984ej,Bando:1987br,Nagahiro:2008cv}
\begin{eqnarray} \label{Eq: lagrangian}
 \mathcal{L}_{PPV} &=&-ig~ \langle V_\mu[P,\partial^\mu P]\rangle,\\
 \mathcal{L}_{VVP} &=&\frac{G'}{\sqrt{2}}~\epsilon^{\mu\nu\alpha\beta}\langle\partial_\mu V_\nu \partial_\alpha V_\beta P\rangle, \label{Eq:VVP}\\
 \mathcal{L}_{VVV}&=&ig ~\langle (V_\mu\partial^\nu V^\mu-\partial^\nu V_\mu V^\mu) V_\nu\rangle,
\end{eqnarray}
with $G'=\frac{3g'^2}{4\pi^2f_{\pi}}$, $g'=-\frac{G_{V}m_{\rho}}{\sqrt{2}{f_{\pi}}^2}$, $G_V\simeq 55$ MeV and $f_\pi=93$ MeV and the coupling constant $g=M_V/{2f_{\pi}}$, $M_V\simeq 800$ MeV~\cite{Nagahiro:2008cv}.
The $P$ and $V$ are the pseudoscalar and vector matrices under SU(5) symmetries as
\begin{equation}
{{P}} =
\left(
\begin{array}{ccccc}
 \frac{\sqrt{3}\pi^0+\sqrt{2}\eta+\eta'}{\sqrt{6}} & \pi^+ & K^+ & \bar{D}^0&B^+ \\
\pi^- &  \frac{-\sqrt{3}\pi^0+\sqrt{2}\eta+\eta'}{\sqrt{6}}  & K^0 & D^- &B^0\\
K^- & \bar{K}^0 & \frac{-\eta+\sqrt{2}\eta'}{\sqrt{3}} & D_s^- &B_s^0\\
D^0 & D^+ & D_s^+ & \eta_c&B_c^+\\
B^-&\bar{B^0}&\bar{B_s^0}&B_c^-&\eta_b\\
\end{array}
\right)\, ,\label{Pmatrix5}
\end{equation}
and
\begin{equation}
{V} =
\left(
\begin{array}{ccccc}
 \frac{\rho^0+\omega}{\sqrt{2}} & \rho^+ & K^{* +} & \bar{D}^{* 0} &B^{*+}\\
\rho^- & \frac{-\rho^0+\omega}{\sqrt{2}}
 & K^{* 0} & {D}^{* -} &B^{*0}\\
K^{* -} & \bar{K}^{* 0} & \phi & D_s^{* -} &B_s^{*0}\\
D^{* 0} & D^{* +} & D_s^{* +} & J/\psi&B_c^{*+}\\
B^{*-}&\bar{B}^{*0}&\bar{B_s}^{*0}&B_c^{*-}&\Upsilon\\
\end{array}
\right)\, .\label{Vmatrix5}
\end{equation}
Generally speaking, the SU(5) symmetry breaks seriously in the charmed and bottom sectors. However, in the current work the heavy mesons are always taken as the constituent particles of the system and almost on shell, which will reduce the effect of the symmetry breaking.

With the above Lagrangians for the vertices, the potential kernel can be constructed in the one boson-exchange model with the help of the standard Feynman rule as~\cite{He:2019ify}
\begin{align}
{\cal V}_{{P}}&=I_{{P}}\Gamma_1\Gamma_2 P_{{P}}f_{P}^2(q^2)\equiv I_{{P}} P_{{P}}f_{P}^2(q^2) \tilde{\cal V}_{{P}},\\
{\cal V}_{{V}}&=I_{{V}}\Gamma_{1\mu}\Gamma_{2\nu}  P^{\mu\nu}_{{V}}f_{V}^2(q^2)
\equiv I_{{V}} P_{{V}}f_{V}^2(q^2)\tilde{\cal V}_{{V}},\label{V}
\end{align}
where the $\Gamma_1$ and $\Gamma_2$ are for the upper and lower vertices of the one-boson exchange Feynman diagram, respectively. The form factor $f_{P,V}$ will be given later. The propagators are defined as
\begin{equation}
P_{{P},{V}}= \frac{i}{q^2-m_{{P}}^2},\ \
P^{\mu\nu}_{V}=i\frac{-g^{\mu\nu}+q^\mu q^\nu/m^2_{{V}}}{q^2-m_{V}^2},
\end{equation}
where  $q$ is the momentum of  exchanged meson and $m_{{P}}$ and $m_{{V}}$ represent the masses of the  exchanged pseudoscalar and vector mesons, respctviely. The $I_{{P},{V}}$ is the flavor factors for certain meson exchange which can be derived with the Lagrangians in Eq.~(\ref{Eq: lagrangian}) and the matrices in Eqs.~(\ref{Pmatrix5}) and (\ref{Vmatrix5}). The explicit values are listed in Table~\ref{flavor factor}.

\renewcommand\tabcolsep{0.2cm}
\renewcommand{\arraystretch}{1.6}
\begin{table}[h!]
\caption{The flavor factors for certain meson exchanges of certain interaction. The values in brackets are for the case of $I=1$ if the values are different from these of $I=0$. The vertex for three pseudoscalar  mesons should be forbidden. \label{flavor factor}}
\begin{tabular}{cccccc}\toprule[2pt]\hline
& $\pi$&$\eta$&$\eta'$ & $\rho$ & $\omega$ \\\hline
$D^{(*)}K^{(*)}/\bar{B}^{(*)}K^{(*)}$&$3/2[-1/2]$&$0$&$1/2$ & $-3/2[1/2]$ &$-1/2$\\
 $\bar{D}^{(*)}K^{(*)}/B^{(*)}K^{(*)}$&$-3/2[1/2]$ &$0$&$1/2$ &$-3/2[1/2]$  &$1/2$\\\hline
\bottomrule[2pt]
\end{tabular}
\end{table}

With the above information, the explicit potential kernels can be constructed explicitly as,
\begin{align}
\label{vvpp}
i\tilde{\cal V}_{V}^{PP}&= -g^2~(k'_1+k_1)\cdot(k'_2+k_2),\\
\label{vvpv}
i\mathcal{V}_{V}^{PV}&=-{g^2}~\Big[(k'_1+k_1)\cdot\epsilon'_2
~(q+k'_2)\cdot\epsilon_2-(k'_1+k_1)\cdot\epsilon_2~\nonumber\\
&\times(q-k_2)\cdot\epsilon'_2
-(k'_1+k_1)\cdot(k'_2+k_2)~\epsilon'_2\cdot\epsilon_2\Big],
\\
\label{vvvp}
i\tilde{\cal V}_{V}^{VP}&=-{g^2}~\Big[(k'_2+k_2)\cdot\epsilon_1
~(q+k_1)\cdot\epsilon'_1-(k'_2+k_2)\cdot\epsilon'_1\nonumber\\
&\times(q-k_1')~\epsilon_1-(k'_2+k_2)
\cdot(k'_1+k_1)~\epsilon'_1\cdot\epsilon_1\Big],
\\
i\tilde{\cal V}_{V}^{VV}&=-g^2\Big[(q-k'_1)\cdot \epsilon_1~\epsilon'_1\cdot \epsilon_2~(q-k_2)\cdot \epsilon'_2~\nonumber\\
&-(q-k'_1)\cdot \epsilon_1\epsilon'_1\cdot \epsilon'_2~(q+k'_2)\cdot \epsilon_2\nonumber\\
&+(q-k'_1)\cdot \epsilon_1~\epsilon'_1\cdot (k_2+k'_2)~\epsilon_2\cdot \epsilon'_2\nonumber\\
&-(q+k_1)\cdot \epsilon'_1~\epsilon_1\cdot \epsilon_2~(q-k_2)\cdot \epsilon'_2\nonumber\\
&+(q+k_1)\cdot \epsilon'_1~\epsilon_1\cdot \epsilon'_2~(q+k'_2)\cdot \epsilon_2\nonumber\\
&-(q+k_1)\cdot \epsilon'_1~\epsilon_1\cdot (k_2+k'_2)~\epsilon_2\cdot \epsilon'_2\nonumber\\
&+\epsilon_1\cdot \epsilon'_1~(k_1+k'_1)\cdot \epsilon_2~(q-k_2)\cdot \epsilon'_2\nonumber\\
&-\epsilon_1\cdot \epsilon'_1~(k_1+k'_1)\cdot \epsilon'_2~(q+k'_2)\cdot \epsilon_2\nonumber\\
&+\epsilon_1\cdot \epsilon'_1~(k_1+k'_1)\cdot(k_2+k'_2)~\epsilon_2\cdot \epsilon'_2\Big],
\\
\label{vvvv}
i\tilde{\cal V}_{P}^{VV}
&=\frac{G'^{2}}{2}~\epsilon_{\mu\nu\alpha\beta
} k_{1}^{\mu}~\epsilon_{1}^{\nu}~k'^{\alpha}_{1}\cdot\epsilon'^{\beta}_{1}~\epsilon_{\mu'\nu'\alpha'\beta'}
k'^{\mu'}_{2}~\epsilon'^{\nu'}_{2}~k_{2}^{\alpha'}~\epsilon_{2}^{\beta'},
\end{align}
where $k_{1, 2}$ and $k'_{1,2}$  are the momenta for the initial and final particle 1 or 2, and the momenta for exchanged meson are defined as $q=k_2'-k_2$.  The $\epsilon$ is the polarized vector for  vector meson. The superscript $VV$ and subscript $P$ of $\mathcal{V}_{P}^{VV}$ means that  two vector constituent mesons  interact by pseudoscalar exchange. Others are defined analogously.

The Bethe-Salpeter equation is  widely used to treat two-body scattering. The potentials obtained above can be taken as the ladder approximation kernel of the Bethe-Salpeter equation, which  describes the interaction well.  In order to reduce the 4-dimensional Bethe-Salpeter equation to a 3-dimensional equation, we adopt the covariant spectator approximation, which  keeps the unitary and covariance of the equation~\cite{Gross:1991pm}. In such treatment, one of the constituent particles, usually the heavier one, is put on shell, which leads to a reduced propagator for two constituent particles in the center-of-mass frame as~\cite{He:2014mja,He:2011ed}
\begin{eqnarray}
	G_0&=&\frac{\delta^+(p''^{~2}_h-m_h^{2})}{p''^{~2}_l-m_l^{2}}\nonumber\\
          &=&\frac{\delta^+(p''^{0}_h-E_h({\rm p}''))}{2E_h({\rm p''})[(W-E_h({\rm
p}''))^2-E_l^{2}({\rm p}'')]}.
\end{eqnarray}
As required by the spectator approximation, the heavier particle  (marked with $h$) satisfies $p''^0_h=E_{h}({\rm p}'')=\sqrt{
m_{h}^{~2}+\rm p''^2}$. The $p''^0_l$ for the lighter particle (marked as $l$) is then $W-E_{h}({\rm p}'')$. Here and hereafter, we define the value of the momentum  in center-of-mass frame as ${\rm p}=|{\bm p}|$.

After the covariant spectator approximation, the 3-dimensional Bethe-Saltpeter equation can be reduced to a 1-dimensional equation with fixed spin parity $J^P$ by partial-wave decomposition~\cite{He:2014mja},
\begin{eqnarray}
i{\cal M}^{J^P}_{\lambda'\lambda}({\rm p}',{\rm p})
&=&i{\cal V}^{J^P}_{\lambda',\lambda}({\rm p}',{\rm
p})+\sum_{\lambda''}\int\frac{{\rm
p}''^2d{\rm p}''}{(2\pi)^3}\nonumber\\
&\cdot&
i{\cal V}^{J^P}_{\lambda'\lambda''}({\rm p}',{\rm p}'')
G_0({\rm p}'')i{\cal M}^{J^P}_{\lambda''\lambda}({\rm p}'',{\rm
p}),\quad\quad \label{Eq: BS_PWA}
\end{eqnarray}
where the sum extends only over non-negative helicity $\lambda''$.

The partial-wave potential can be calculated from the potential kernel obtained in Eqs.~(\ref{vvpp})-(\ref{vvvv}) as
\begin{eqnarray}
{\cal V}_{\lambda'\lambda}^{J^P}({\rm p}',{\rm p})
&=&2\pi\int d\cos\theta
~[d^{J}_{\lambda\lambda'}(\theta)
{\cal V}_{\lambda'\lambda}({\bm p}',{\bm p})\nonumber\\
&+&\eta d^{J}_{-\lambda\lambda'}(\theta)
{\cal V}_{\lambda'-\lambda}({\bm p}',{\bm p})],
\end{eqnarray}
where $\eta=PP_1P_2(-1)^{J-J_1-J_2}$ with $P$ and $J$ being parity and spin for the system. The initial and final relative momenta are chosen as ${\bm p}=(0,0,{\rm p})$ and ${\bm p}'=({\rm p}'\sin\theta,0,{\rm p}'\cos\theta)$. The $d^J_{\lambda\lambda'}(\theta)$ is the Wigner $d$-matrix.

To compensate the off-shell effect and the nonpointlike nature of the hadrons involved,  a form factor is introduced as $f(q^2)=e^{-(k^2-m^2)^2/\Lambda^2}$ to all off-shell mesons,
where $k$ and $m$ are the momentum and mass of  meson. The cutoffs in the form factors for the exchanged mesons and cutoff for constituent mesons are chosen as the same for simplification.

The partial-wave qBSE is a one-dimensional integral equation, which can be solved by discretizing the momenta with the Gauss quadrature.  It leads to a matrix equation of a form
$M=V+VGM$~\cite{He:2014mja}. The molecular state corresponds to the pole of the amplitude, which can be obtained by varying $z$ to satisfy $|1-V(z)G(z)|=0$
with  $z=E_R-i\Gamma/2$ being the exact position of the bound state.

 \section{Numerical results}\label{Sec: results}

With previous preparations, we can perform calculation about the systems considered. The systems $D^{(*)}K^{(*)}$ and $\bar{B}^{(*)}K^{(*)}$  with quark configuration $[Q\bar{q}][q\bar{s}]$ will be studied first. The isoscalar bound state from such systems corresponds to experimentally observed $D^*_{s0}(2317)$ and $D_{s1}(2460)$.  Then, we will make  study of the bound states of systems $\bar{D}^{(*)}K^{(*)}$ and ${B}^{(*)}K^{(*)}$  with quark configuration $[\bar{Q}q][q\bar{s}]$  as the $X_0(2900)$, which cannot correspond to a meson in conventional quark model. The single-channel results will be given first, and the coupled-channel effect will be discussed at the end of this section with a coupled-channel calculation. 

\subsection{The molecular states from the systems $D^{(*)}K^{(*)}$ and $\bar{B}^{(*)}K^{(*)}$   with quark configuration $[Q\bar{q}][q\bar{s}]$}

For the systems $D^{(*)}K^{(*)}$ and $\bar{B}^{(*)}K^{(*)}$, we will consider all possible $S$-wave interactions to search for the near-threshold bound states, which involve spin parities $J^P=0^+$ for systems $DK$ and $\bar{B}K$ , $1^+$ for systems $D^*K$, $\bar{B}^*K$,  $DK^*$, and $\bar{B}K^*$, and $(0,1,2)^+$ for systems $D^*K^*$ and $\bar{B}^*K^*$. The isospins of the heavy meson and strange meson produce both isoscalar and isovector states. Hence, 24 states will be considered in the calculation. In the current model, the only free parameter is  the cutoff $\Lambda$. We vary it  in a range of 0.5-5 GeV to search for the bound state with binding energy smaller than 50~MeV.  The results of these states are listed in  Table~\ref{qqbar}.

\renewcommand\tabcolsep{0.32cm}
\renewcommand{\arraystretch}{1.}
\begin{table}[hpbt!]
\caption{The binding energies of the bound states from the   interactions of $D^{(*)}K^{(*)}$ and $\bar{B}^{(*)}K^{(*)}$ with some selected values of cutoff $\Lambda$. The ``$--$" means that no bound state is found in the considered range of the cutoff.
The units of the cutoff $\Lambda$ and binding energy $E_B$ are GeV and MeV, respectively.\label{qqbar}}
\begin{tabular}{cccccccc}
\toprule[2pt]\hline
\multicolumn{4}{c}{$DK$} &\multicolumn{4}{c}{$\bar{B}K$}\\\cmidrule(lr){1-4}\cmidrule(lr){5-8}
\multicolumn{2}{c}{$0(0^+)$} &\multicolumn{2}{c}{$1(0^+)$}&
\multicolumn{2}{c}{$0(0^+)$} &\multicolumn{2}{c}{$1(0^+)$}\\\cmidrule(lr){1-2} \cmidrule(lr){3-4}\cmidrule(lr){5-6} \cmidrule(lr){7-8}
$\Lambda$ &$E_B$&$\Lambda$ &$E_B$&$\Lambda$ &$E_B$ &$\Lambda$ &$E_B$\\
$1.30$&$1$   &$--$&$--$   &$1.20$ &$1$   &$--$&$--$ \\
$1.40$&$6$ &$--$&$--$   &$1.40$ &$11$  &$--$&$--$  \\
$1.80$&$39$ &$--$&$--$   &$1.60$ &$26$ &$--$&$--$   \\
\toprule[1pt]
\multicolumn{4}{c}{$D^*K$} &\multicolumn{4}{c}{$\bar{B}^*K$}\\\cmidrule(lr){1-4}\cmidrule(lr){5-8}
\multicolumn{2}{c}{$0(1^+)$} &\multicolumn{2}{c}{$1(1^+)$}&
\multicolumn{2}{c}{$0(1^+)$} &\multicolumn{2}{c}{$1(1^+)$}\\\cmidrule(lr){1-2} \cmidrule(lr){3-4}\cmidrule(lr){5-6} \cmidrule(lr){7-8}
$\Lambda$ &$E_B$&$\Lambda$ &$E_B$&$\Lambda$ &$E_B$ &$\Lambda$ &$E_B$\\
$1.30$&$1$   &$--$&$--$   &$1.21$ &$1$   &$--$&$--$ \\
$1.40$&$5$ &$--$&$--$    &$1.29$ &$4$  &$--$&$--$  \\
$1.80$&$36$ &$--$&$--$   &$1.40$ &$10$ &$--$&$--$   \\
\toprule[1pt]
\multicolumn{4}{c}{$DK^*$} &\multicolumn{4}{c}{$\bar{B}K^*$}\\\cmidrule(lr){1-4}\cmidrule(lr){5-8}
\multicolumn{2}{c}{$0(1^+)$} &\multicolumn{2}{c}{$1(1^+)$}&
\multicolumn{2}{c}{$0(1^+)$} &\multicolumn{2}{c}{$1(1^+)$}\\\cmidrule(lr){1-2} \cmidrule(lr){3-4}\cmidrule(lr){5-6} \cmidrule(lr){7-8}
$\Lambda$ &$E_B$&$\Lambda$ &$E_B$&$\Lambda$ &$E_B$ &$\Lambda$ &$E_B$\\
$1.10$&$1$   &$--$&$--$   &$1.00$ &$1$   &$--$&$--$ \\
$1.40$&$23$ &$--$&$--$   &$1.40$ &$36$  &$--$&$--$  \\
$1.60$&$48$ &$--$&$--$   &$1.50$ &$48$ &$--$&$--$   \\
\toprule[1pt]
\multicolumn{4}{c}{$D^*K^*$} &\multicolumn{4}{c}{$\bar{B}^*K^*$}\\\cmidrule(lr){1-4}\cmidrule(lr){5-8}
\multicolumn{2}{c}{$0(0^+)$} &\multicolumn{2}{c}{$1(0^+)$}&
\multicolumn{2}{c}{$0(0^+)$} &\multicolumn{2}{c}{$1(0^+)$}\\\cmidrule(lr){1-2} \cmidrule(lr){3-4}\cmidrule(lr){5-6} \cmidrule(lr){7-8}
$\Lambda$ &$E_B$&$\Lambda$ &$E_B$&$\Lambda$ &$E_B$ &$\Lambda$ &$E_B$\\
$1.10$&$1$   &$--$&$--$   &$1.00$ &$1$   &$--$&$--$ \\
$1.40$&$6$ &$--$&$--$   &$1.40$ &$35$  &$--$&$--$  \\
$1.60$&$45$ &$--$&$--$   &$1.50$ &$47$ &$--$&$--$   \\
\multicolumn{2}{c}{$0(1^+)$} &\multicolumn{2}{c}{$1(1^+)$}&
\multicolumn{2}{c}{$0(1^+)$} &\multicolumn{2}{c}{$1(1^+)$}\\\cmidrule(lr){1-2} \cmidrule(lr){3-4}\cmidrule(lr){5-6} \cmidrule(lr){7-8}
$\Lambda$ &$E_B$&$\Lambda$ &$E_B$&$\Lambda$ &$E_B$ &$\Lambda$ &$E_B$\\
$1.10$&$1$   &$--$&$--$   &$1.00$ &$1$   &$--$&$--$ \\
$1.40$&$22$ &$--$&$--$   &$1.40$ &$35$  &$--$&$--$  \\
$1.60$&$46$ &$--$&$--$   &$1.50$ &$47$ &$--$&$--$   \\
\multicolumn{2}{c}{$0(2^+)$} &\multicolumn{2}{c}{$1(2^+)$}&
\multicolumn{2}{c}{$0(2^+)$} &\multicolumn{2}{c}{$1(2^+)$}\\\cmidrule(lr){1-2} \cmidrule(lr){3-4}\cmidrule(lr){5-6} \cmidrule(lr){7-8}
$\Lambda$ &$E_B$&$\Lambda$ &$E_B$&$\Lambda$ &$E_B$ &$\Lambda$ &$E_B$\\
$1.10$&$1$   &$--$&$--$   &$1.10$ &$1$   &$--$&$--$ \\
$1.40$&$24$ &$--$&$--$   &$1.40$ &$24$  &$--$&$--$  \\
$1.60$&$50$ &$--$&$--$   &$1.50$ &$36$ &$--$&$--$   \\
\hline
\bottomrule[2pt]
\end{tabular}
\end{table}

\begin{figure}[h!]
\centering
\includegraphics[scale=0.65,bb=35 135 371 520,clip]{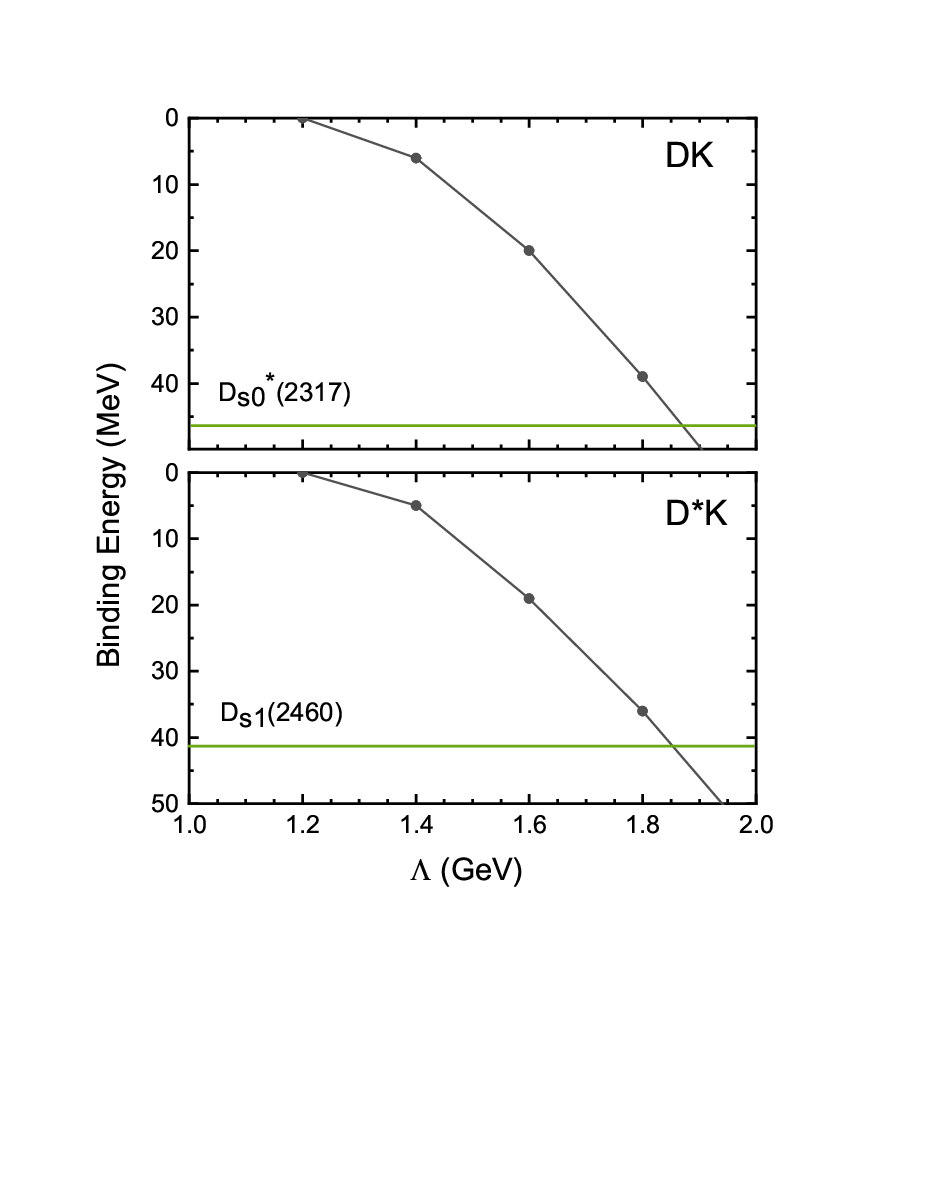}
\caption{The binding energy $E_B$ of the  bound states $DK(0^+)$ (upper panel) and $D^*K(1^+)$ (lower panel) with the variation of cutoff $\Lambda$. Here  $E_B=M_{th}-W$, with  $M_{th}$ and W being the threshold and mass of the state. The green lines are the experimental mass of the $D^*_{s0}(2317)$ and $D_{s1}(2460)$.  We would like to remind that the experimental mass uncertainties of these two states are very small. }
\label{DK3}
\end{figure}

For the $DK$ system with the lowest threshold amongst the interactions considered in the current work, an isoscalar bound state is produced  with spin parity $J^P=0^+$ at a cutoff of about 1.3~GeV and its binding energy increases to 39~MeV with the increase of the cutoff $\Lambda$ to 1.8~GeV.  If we replace a $D$ meson by the vector meson $D^*$, an isoscalar $D^*K$ bound state is also produced with $1^+$ at a cutoff of about 1.3 GeV analogously to the $DK$ system.  We illustrate the variation of binding energies of these two states with the function of the cutoff $\Lambda$ in Fig.~\ref{DK3}, and compare with the experimental masses of the charmed-strange mesons $D^*_{s0}(2317)$ and $D_{s1}(2460)$. The isospin spin parities of $D^*_{s0}(2317)$ and $D_{s1}(2460)$ are $0(0^+)$ and $0(1^+)$, which just correspond to the quantum numbers  of  $S$-wave interactions of $DK$ and $DK^*$, respectively. It is  natural to relate the $D^*_{s0}(2317)$ and $D_{s1}(2460)$ to the  bound states $DK(0^+)$ and $D^*K(1^+)$, respectively. As shown in the figure, the tendencies of the curves for two states are analogous, and the experimental masses can be reproduced at a cutoff about 1.8 GeV for both states.

For the $DK^*$ system, there also exists only one isoscalar S-wave state, which is bound at a cutoff of about 1.1 GeV, a little smaller than the cutoffs required for the systems $DK$ and $D^*K$.
In the above three systems, the existence of a pseudoscalar meson only permits one isoscalar $S$-wave state.  For the system with  two vector mesons, there are three isoscalar $S$-wave states. The calculation suggests that bound states can be produced for all three isoscalar states with spin parities $0^+$, $1^+$ and $2^+$. The binding energies of these three states are very close. To further distinguish them, we may need partial-wave analysis. Therefore, the bound states are found from all  six $S$-wave isoscalar interactions of  charm meson $D^{(*)}$ and strange meson $K^{(*)}$.

In the right part of Table~\ref{qqbar}, we present the $\bar{B}^{(*)}K^{(*)}$ systems with an antibottom meson. One can find that the results in this sector are very analogous to those in the charmed sector.  It is easy to understand because the potential in the bottom sector has the same form as that in the charmed sector. And, we can also notice that smaller cutoff in the bottom sector is required to produce a bound state. It is due to  larger reduced mass of systems in bottom sector than the charmed system.

\begin{figure}[h!]
\centering
\includegraphics[scale=0.7,bb=40 170 420 421,clip]{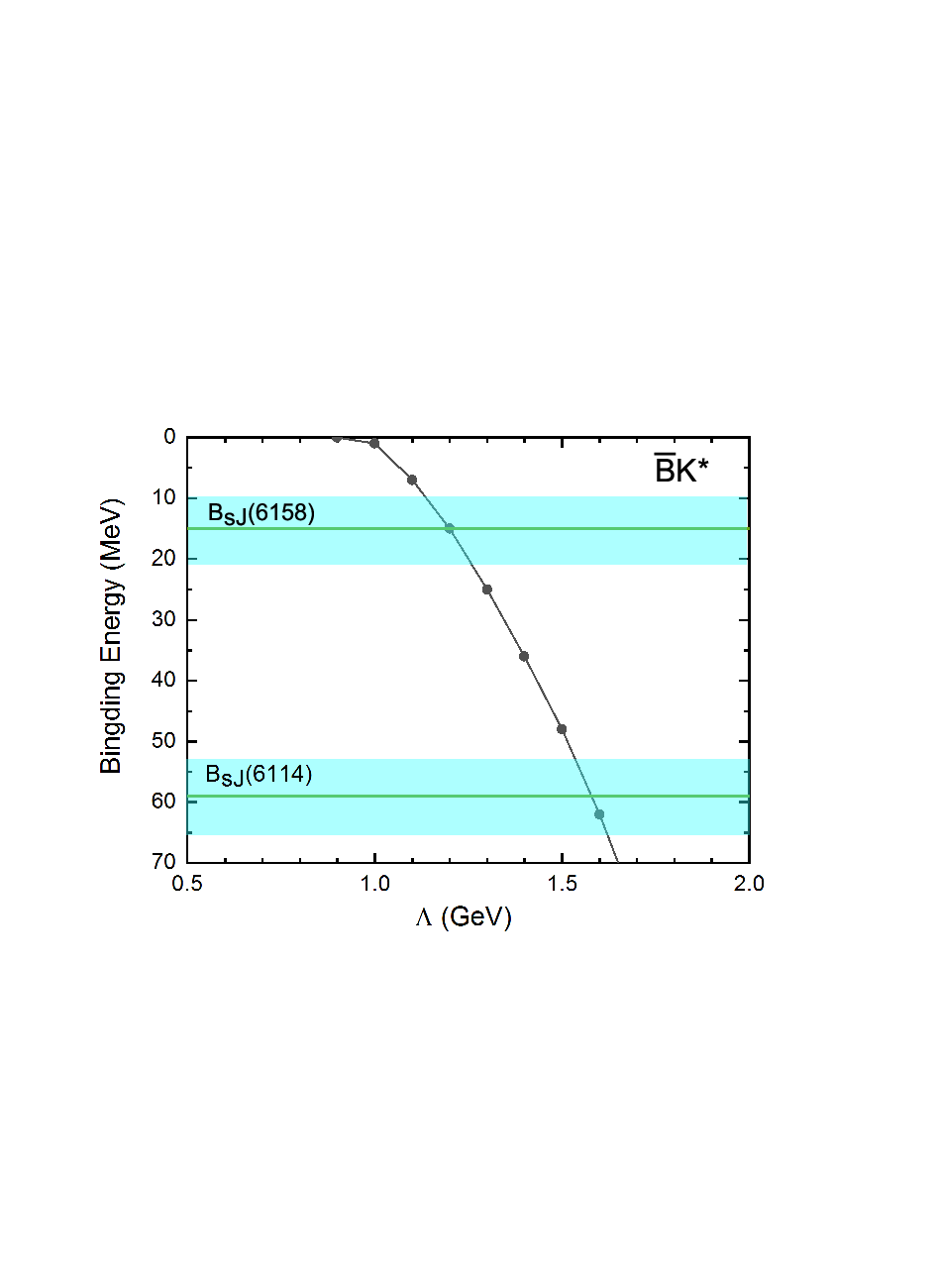}
\caption{The binding energy $E_B$ of $\bar{B}K^*(1^+)$ bound state with the variation of cutoff $\Lambda$. Here the $E_B=M_{th}-W$, the $M_{th}$ and W being the threshold and mass of the state. The green lines and the
band are the experimental mass and uncertainties of the $B_{sJ}(6158)$ and $B_{sJ}(6114)$.}
\label{BK4}
\end{figure}

Recently, LHCb reported two bottom-strange states $B_{sJ}(6064)$ and $B_{sJ}(6114)$ in $B^+K^-$ decay or  $B_{sJ}(6109)$ and $B_{sJ}(6158)$ in $B^{*+}K^-$ decay. These states are close to the $\bar{B}K^*$ threshold, which is about 6.170~GeV. Hence, it is natural to relate them to the $\bar{B}K^*$  molecular state. From the  $\bar{B}K^*$ interaction, only one isoscalar state with  $1^+$ is produced in $S$ wave.
The result for the $\bar{B}K^*$ state is illustrated in Fig.~\ref{BK4}. If we consider the states  $B_{sJ}(6064)$, $B_{sJ}(6114)$, and $B_{sJ}(6109)$ are much lower than the  $\bar{B}K^*$ threshold,  the $B_{sJ}(6158)$ is the best candidate of a $\bar{B}K^*$ molecular state with spin parity $0(1^+)$.

It is very interesting to notice that only isoscalar states can be produced, while none of  isovector states are  produced in both charmed and bottom sectors. If we recall the flavor factors
listed in Table.~\ref{flavor factor} and the form of potential kernels in Eq.~(\ref{Eq: lagrangian}), the main contribution for the potential kernels is from $\pi$ and $\rho$ meson exchanges, which is completely different for two cases with $I=0$ and $I=1$. It leads to a repulse force in the isovector case if the isoscalar interactions are attractive. Hence, It is no surprise that no isovector bound states can be produced.

\subsection{The molecular states from the systems $\bar{D}^{(*)}K^{(*)}$ and ${B}^{(*)}K^{(*)}$   with quark configuration $[\bar{Q}q][q\bar{s}]$}

Now we turn to the systems with quark configuration $[\bar{Q}q][q\bar{s}]$ by replacing the heavy meson in the systems discussed in the previous subsection by their antiparticle. We also only consider $S$-wave states with both scalar and vector isospins, which leads to 24 states listed in Table~\ref{Qbar}.

\renewcommand\tabcolsep{0.32cm}
\renewcommand{\arraystretch}{1.}
\begin{table}[hpbt!]
\caption{The binding energies of the bound states from  interactions of $\bar{D}^{(*)}K^{(*)}$ and ${B}^{(*)}K^{(*)}$ with some selected values of cutoff $\Lambda$. The ``$--$" means that no bound state is found in the considered range of the cutoff.
The units of the cutoff $\Lambda$ and binding energy $E_B$ are GeV and MeV, respectively.\label{Qbar}}
\begin{tabular}{cccccccc}
\toprule[2pt]\hline
\multicolumn{4}{c}{$\bar{D}K$} &\multicolumn{4}{c}{$BK$}\\\cmidrule(lr){1-4}\cmidrule(lr){5-8}
\multicolumn{2}{c}{$0(0^+)$} &\multicolumn{2}{c}{$1(0^+)$}&
\multicolumn{2}{c}{$0(0^+)$} &\multicolumn{2}{c}{$1(0^+)$}\\\cmidrule(lr){1-2} \cmidrule(lr){3-4}\cmidrule(lr){5-6} \cmidrule(lr){7-8}
$\Lambda$ &$E_B$&$\Lambda$ &$E_B$&$\Lambda$ &$E_B$ &$\Lambda$ &$E_B$\\
$4.0$&$1$   &$--$&$--$   & $4.0$&$0.3$  &$--$&$--$ \\
$4.2$&$2$  &$--$&$--$   & $4.2$&$0.4$  &$--$&$--$ \\
$4.4$&$3$  &$--$&$--$   & $4.4$&$0.6$  &$--$&$--$ \\
\toprule[1pt]
\multicolumn{4}{c}{$\bar{D}^*K$} &\multicolumn{4}{c}{${B}^*K$}\\\cmidrule(lr){1-4}\cmidrule(lr){5-8}
\multicolumn{2}{c}{$0(1^+)$} &\multicolumn{2}{c}{$1(1^+)$}&
\multicolumn{2}{c}{$0(1^+)$} &\multicolumn{2}{c}{$1(1^+)$}\\\cmidrule(lr){1-2} \cmidrule(lr){3-4}\cmidrule(lr){5-6} \cmidrule(lr){7-8}
$\Lambda$ &$E_B$&$\Lambda$ &$E_B$&$\Lambda$ &$E_B$ &$\Lambda$ &$E_B$\\
$--$&$--$ &$--$&$--$   &$--$&$--$   &$--$&$--$ \\
$--$&$--$ &$--$&$--$    &$--$&$--$  &$--$&$--$  \\
$--$&$--$ &$--$&$--$   &$--$&$--$ &$--$&$--$   \\
\toprule[1pt]
\multicolumn{4}{c}{$\bar{D}K^*$} &\multicolumn{4}{c}{${B}K^*$}\\\cmidrule(lr){1-4}\cmidrule(lr){5-8}
\multicolumn{2}{c}{$0(1^+)$} &\multicolumn{2}{c}{$1(1^+)$}&
\multicolumn{2}{c}{$0(1^+)$} &\multicolumn{2}{c}{$1(1^+)$}\\\cmidrule(lr){1-2} \cmidrule(lr){3-4}\cmidrule(lr){5-6} \cmidrule(lr){7-8}
$\Lambda$ &$E_B$&$\Lambda$ &$E_B$&$\Lambda$ &$E_B$ &$\Lambda$ &$E_B$\\
$2.00$&$1$   &$--$&$--$   &$1.60$ &$1$   &$--$&$--$ \\
$2.20$&$5$ &$--$&$--$   &$1.80$ &$4$  &$--$&$--$  \\
$2.40$&$11$ &$--$&$--$   &$2.00$ &$8$ &$--$&$--$   \\
\toprule[1pt]
\multicolumn{4}{c}{$\bar{D}^*K^*$} &\multicolumn{4}{c}{${B}^*K^*$}\\\cmidrule(lr){1-4}\cmidrule(lr){5-8}
\multicolumn{2}{c}{$0(0^+)$} &\multicolumn{2}{c}{$1(0^+)$}&
\multicolumn{2}{c}{$0(0^+)$} &\multicolumn{2}{c}{$1(0^+)$}\\\cmidrule(lr){1-2} \cmidrule(lr){3-4}\cmidrule(lr){5-6} \cmidrule(lr){7-8}
$\Lambda$ &$E_B$&$\Lambda$ &$E_B$&$\Lambda$ &$E_B$ &$\Lambda$ &$E_B$\\
$2.00$&$1$   &$--$&$--$   &$1.60$ &$1$   &$--$&$--$ \\
$2.20$&$2$ &$--$&$--$   &$1.80$ &$3$  &$--$&$--$  \\
$2.40$&$5$ &$--$&$--$   &$2.00$ &$7$ &$--$&$--$   \\
\multicolumn{2}{c}{$0(1^+)$} &\multicolumn{2}{c}{$1(1^+)$}&
\multicolumn{2}{c}{$0(1^+)$} &\multicolumn{2}{c}{$1(1^+)$}\\\cmidrule(lr){1-2} \cmidrule(lr){3-4}\cmidrule(lr){5-6} \cmidrule(lr){7-8}
$\Lambda$ &$E_B$&$\Lambda$ &$E_B$&$\Lambda$ &$E_B$ &$\Lambda$ &$E_B$\\
$2.00$&$1$   &$--$&$--$   &$1.60$ &$1$   &$--$&$--$ \\
$2.20$&$4$ &$--$&$--$   &$1.80$ &$3$  &$--$&$--$  \\
$2.40$&$9$ &$--$&$--$   &$2.00$ &$7$ &$--$&$--$   \\
\multicolumn{2}{c}{$0(2^+)$} &\multicolumn{2}{c}{$1(2^+)$}&
\multicolumn{2}{c}{$0(2^+)$} &\multicolumn{2}{c}{$1(2^+)$}\\\cmidrule(lr){1-2} \cmidrule(lr){3-4}\cmidrule(lr){5-6} \cmidrule(lr){7-8}
$\Lambda$ &$E_B$&$\Lambda$ &$E_B$&$\Lambda$ &$E_B$ &$\Lambda$ &$E_B$\\
$2.00$&$1$   &$--$&$--$   &$1.60$ &$1$   &$--$&$--$ \\
$2.20$&$4$ &$--$&$--$   &$1.80$ &$4$  &$--$&$--$  \\
$2.40$&$11$ &$--$&$--$   &$2.00$ &$8$ &$--$&$--$   \\
\hline
\bottomrule[2pt]
\end{tabular}
\end{table}

For the $\bar{D}K$ interaction, a very large cutoff, about 4.0 GeV, is required to produce its isoscalar $S$-wave state with $0(0^+)$. Since such value is very large itself and compared with the values for other states, we do not suggest  existence of this state.  For the $\bar{D}^*K$ system, no bound state can be produced in the range of the cutoff considered in the current work.  For the $\bar{D}K^*$ system, an isoscalar state with $J^P=1^+$ appears at a cutoff of about 2 GeV.

\begin{figure}[h!]
\centering
\includegraphics[scale=0.7,bb=40 140 400 380,clip]{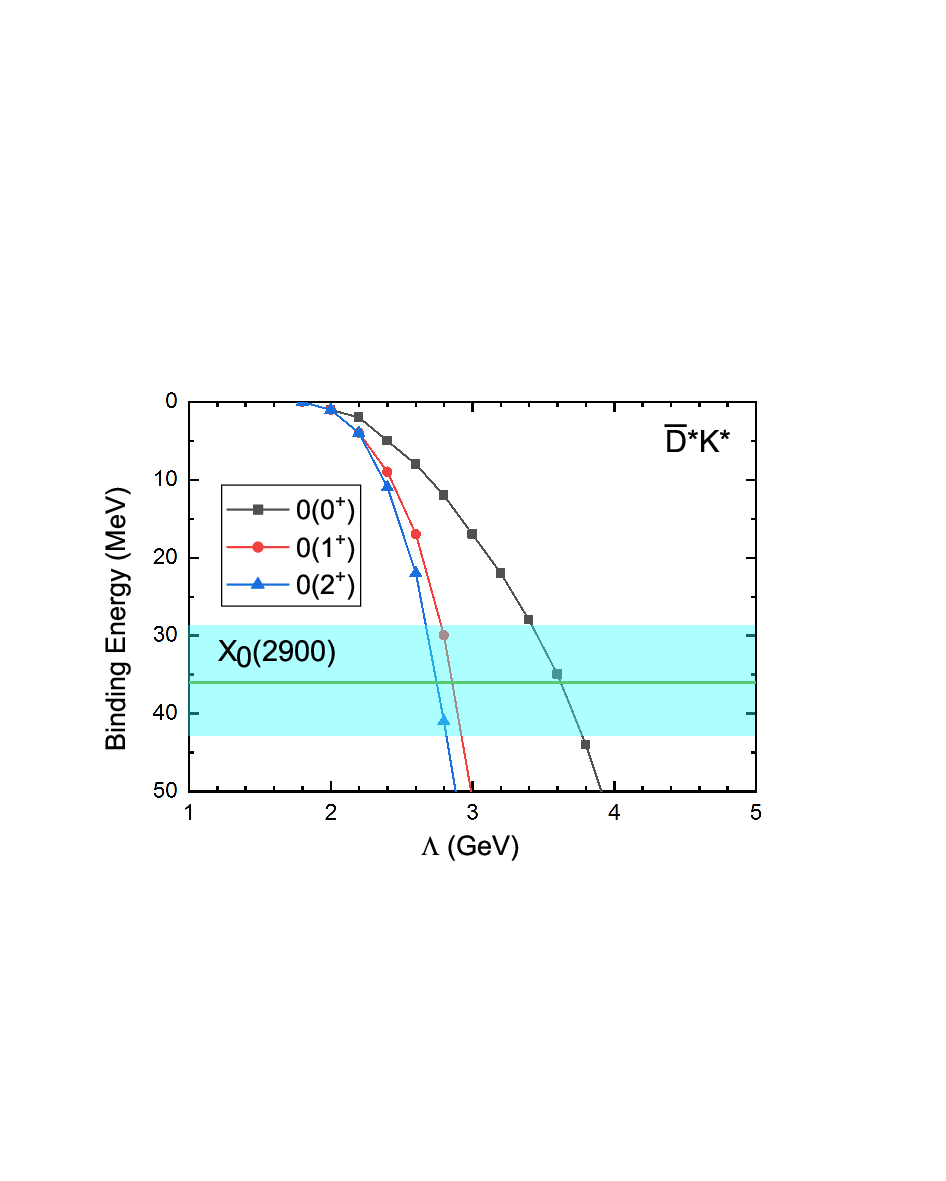}
\caption{The binding energy $E_B$ of $\bar{D}^*K^*$ bound states with the variation of the $\Lambda$. Here the $E_B=M_{th}-W$, the $M_{th}$ and W being the threshold and mass of the state. The green line and the
band are the experimental mass and uncertainty of the $X(2900)$.}
\label{X2900}
\end{figure}

For the systems of two vector mesons,  we have three total spins $J=0$, 1, and 2.  In the calculation, all isoscalar $S$-wave states are produced from the $\bar{D}^*K^*$ interaction. All three states appear at a cutoff $\Lambda$ of about 2~GeV, and variation of the binding energies with cutoff is relatively slower compared with systems with quark configuration $[Q\bar{q}][q\bar{s}]$.  Obviously, as discussed in the literature~\cite{Molina:2010tx,Xue:2020vtq,Liu:2020nil,Chen:2020aos,Agaev:2020nrc,Huang:2020ptc,Mutuk:2020igv,Molina:2020hde,He:2020btl,Xiao:2020ltm}, the $X_0(2900)$ reported by LHCb Collaboration can be related to these states. Considering the spin of $X_0(2900)$, it should correspond to the $0^+$ state of  $\bar{D}^*K^*$ interaction. We illustrate the theoretical binding energy with the variation of the cutoff and the experimental result about the $X_0(2900)$ in Fig.~\ref{X2900}.
The binding energies of the other two bound states in $\bar{D}^*K$ channel with $0(1^+)$ and $0(2^+)$ increase faster with the increase of $\Lambda$. That is to say, with the same cutoff $\Lambda$, the $0(2^+)$ interaction have the strongest attraction. The current results are consistent with the results with the same Lagrangians in the chiral unitary approach~~\cite{Molina:2010tx}. In Ref.~\cite{He:2020btl}, we also obtain three reasonable bound states with the same spin-parity quantum numbers by solving the qBSE with the Lagrangians from heavy quark asymmetry for heavy meson and effective Lagrangians for strange meson. However, order of the binding energies for different spin parties is opposite, which is due to the different Lagrangians adopted.  Such results suggest that with different Lagrangians, we can reach a similar qualitative conclusion, for example, whether there exists a bound state from the interaction, but the explicit qualitative results may be quite different.

In the charmed sector, four states can be produced from the isoscalar $S$-wave interactions with similar cutoff.  If we replacing the $\bar{D}^{(*)}$ meson by the bottom meson $B^{(*)}$, the analogous results can be obtained as listed in Table~\ref{Qbar}. There are also four bound states produced in the bottom sector, but such states appears at a smaller cutoff, about 1.6~GeV. No isovector  states are produced from all systems considered. The discussion in the previous subsection is still established here.

\subsection{Numerical results with coupled-channel calculation}
The states with the same quantum numbers will be mixed to obtain physical states due to the coupled-channel effect. In the following, we will include the coupled-channel effect between the states considered above, and the results  are listed in Table~\ref{Tab: CC}.  Because no isovector state is produced as in the single-channel calculation, we only present the isoscalar cases. With the coupled-channel effect, except the  states with the lowest threshold, the pole positions of other states will deviate from the real axis, and acquire imaginary parts, which correspond to width as $\Gamma=-2 {\rm Im} z$. To compare with the single-channel results, in Table~\ref{Tab: CC} we present the position as  $M_{th}-z$  instead of pole position  $z$, with the $M_{th}$ being the nearest threshold. 

\renewcommand\tabcolsep{0.25cm}
\renewcommand{\arraystretch}{1.}
\begin{table}[h!]
\begin{center}
\caption{ The $M_{th}-z$ of the poles from the doubly heavy coupled-channel interaction. The cutoff $\Lambda$ and $M_{th}-z$ are in units of GeV and MeV, respectively.}
\label{Tab: CC}
\begin{tabular}{ccccccccc}\toprule[2pt]\hline
\multicolumn{3}{c}{$0(0^{+})$}&\multicolumn{4}{c}{$0(1^{+})$}\\\cmidrule(lr){1-3}\cmidrule(lr){4-7}
 $\Lambda$& \multicolumn{2}{c}{$M_{th}-z$} &
  $\Lambda$ &\multicolumn{3}{c}{$M_{th}-z$} \\\cmidrule(lr){2-3}\cmidrule(lr){5-7}
& $D{K}$ & $D^{*}{{K}}^{*}$&
 &$D^{*}{K}$  &$D{K}^{*}$&$D^{*}{K}^{*}$\\
1.2&$1$& $\ \ 0+\ \ 6i$&1.1&$--$& $\ \ 3+\ \ 4i$& $\ \ 1+ 0i$\\
 1.3&$4$& $\ \ 4+10i$&1.2&$--$& $10+\ \ 7i$& $\ \ 6+0i$\\
 1.4&$12$& $10+14i$&1.3&$2$& $21+\ \ 9i$& $13+0i$\\
 1.6&$35$& $29+24i$&1.4&$7$& $34+12i$& $21+0i$\\
\multicolumn{3}{c}{$0(0^{+})$}&\multicolumn{4}{c}{$0(1^{+})$}\\\cmidrule(lr){1-3}\cmidrule(lr){4-7}
 $\Lambda$& \multicolumn{2}{c}{$M_{th}-z$} &
  $\Lambda$ &\multicolumn{3}{c}{$M_{th}-z$} \\\cmidrule(lr){2-3}\cmidrule(lr){5-7}
 & $\bar{B}{K}$ & $\bar{B}^{*}{{K}}^{*}$&
 &$\bar{B}^{*}{K}$  &$\bar{B}{K}^{*}$&$\bar{B}^{*}{K}^{*}$\\
 1.1&$--$& $\ \ 6+1i$&1.0&$--$& $\ \ 1+0i$& $\ \ 6+0i$\\
1.2&$1$& $13+2i$&1.1&$--$& $\ \ 7+1i$& $14+0i$\\
 1.3&$5$& $22+3i$&1.2&$1$& $15+2i$& $24+0i$\\
1.4&$12$& $33+4i$&1.3&$4$& $25+2i$& $35+0i$\\
1.5&$20$& $45+5i$&1.4&$11$& $50+3i$& $47+0i$\\

\multicolumn{3}{c}{$0(0^{+})$}&\multicolumn{4}{c}{$0(1^{+})$}\\\cmidrule(lr){1-3}\cmidrule(lr){4-7}
 $\Lambda$& \multicolumn{2}{c}{$M_{th}-z$} &
  $\Lambda$ &\multicolumn{3}{c}{$M_{th}-z$} \\\cmidrule(lr){2-3}\cmidrule(lr){5-7}
& $\bar{D}{K}$ & $\bar{D}^{*}{{K}}^{*}$&
 &$\bar{D}^{*}{K}$  &$\bar{D}{K}^{*}$&$\bar{D}^{*}{K}^{*}$\\
 2.5&$--$& $2+10i$&1.9&$--$& $\ \ 5+\ \ 5i$& $--$\\
2.7&$1$& $10+42i$&2.0&$--$& $\ \ 9+\ \ 7i$& $1+0i$\\
 2.8&$2$& $16+54i$&2.1&$--$& $14+\ \ 9i$& $2+0i$\\
2.9&$4$& $23+60i$&2.2&$--$& $21+12i$& $4+0i$\\
\multicolumn{3}{c}{$0(0^{+})$}&\multicolumn{4}{c}{$0(1^{+})$}\\\cmidrule(lr){1-3}\cmidrule(lr){4-7}
 $\Lambda$& \multicolumn{2}{c}{$M_{th}-z$} &
  $\Lambda$ &\multicolumn{3}{c}{$M_{th}-z$} \\\cmidrule(lr){2-3}\cmidrule(lr){5-7}
& $B{K}$ & $B^{*}{{K}}^{*}$&
 &$B^{*}{K}$  &$B{K}^{*}$&$B^{*}{K}^{*}$\\
 1.5&$--$& $\ \ 0+0i$&1.5&$--$& $--$& $--$\\
1.8&$--$& $\ \ 3+1i$&1.8&$--$& $\ \ 4+1i$& $\ \ 3+0i$\\
2.0&$--$& $\ \ 7+2i$&2.1&$--$& $12+1i$& $10+0i$\\
2.2&$--$& $13+3i$&2.4&$--$& $24+2i$& $20+0i$\\

\hline
\bottomrule[2pt]
\end{tabular}
\end{center}
\end{table}

For  a system composed of a charmed meson and a strange meson with the  $0^+$ spin parity, only two channels $DK$ and $D^*K^*$ can couple to each other. In  Table \ref{qqbar}, the bound states can be found at cutoffs about  1.3 and 1.1~MeV from the $DK$ and $D^*K^*$ interactions, respectively. After including couple-channel effect,  the poles appear at a cutoff about 1.2~MeV, and the higher state acquires a width,  about 20 MeV at cutoff of 1.3 GeV.  For the $1^+$  spin parity, three channels, $D^*K$, $DK^*$, and $D^*K^*$, can couple together. In the single-channel calculation, the $1^+$ bound states can be found at  cutoffs 1.3, 1.1, and 1.1~MeV from these three interactions, respectively. With couplings, we can still find three poles at almost the same cutoff and notice that the state under the $DK^*$ threshold has a  broad width, which represents the strong coupling to the $DK$ channel. The width of $D^*K^*$ state is very small, which represents that the coupling between $D^*K^*$ and the other two channels is very small. It may be from  suppression of vertex $VVP$ in Eq.~(\ref{Eq:VVP}), which is  inevitable in all possible diagrams of these couplings.  For the systems with an antibottom meson and a strange meson, a similar pattern can be found as in the charmed sector. 

For a system with an anticharmed meson and strange meson,  a scalar state appears near $\bar{D}^*K^*$ threshold from the $\bar{D}K-\bar{D}^*K^*$ system at a cutoff about 2.3~MeV, and with the increasing of the cutoff, a state  with a binding energy and width will increase also, which can be related to the $X_0(2900)$ in LHCb.  With the coupled-channel effect, a state appears near $\bar{D}K$ threshold at a cutoff of 2.7 GeV, which is smaller than the cutoff required in the single-channel $\bar{D}K$ interaction, about 4 GeV. The result for the $BK-B^*K^*$ system is analogous while there is still no  state near $BK$ threshold produced due to weakness of interaction.
When considering $\bar{D}^*K-\bar{D}K^*-\bar{D}^*K^*$ interaction,  the bound state of $\bar{D}K^*$ appears at a little smaller cutoff than the one in the single-channel calculation and only it can get a width about 20 MeV, which suggests a large coupled-channel effect. The states near the $\bar{D}^*K^*$ threshold appear at cutoff about 2 GeV, which is close to the single-channel calculation. In the bottom sector, the calculation suggests that smaller cutoff is needed to produced the state near $B^*K^*$ thresholds while larger cutoff is needed to  produced the state near $BK^*$ thresholds than the single-channel calculation.

\section{Summary and discussion}\label{Sec: summary}

In this work, we systematically study the molecular states produced from the interaction of heavy-strange mesons, $D^{(*)}K^{(*)}/\bar{D}^{(*)}K^{(*)}$ and $\bar{B}^{(*)}K^{(*)}/B^{(*)}K^{(*)}$,  in a qBSE approach together with the one-boson exchange model. The potential kernels are constructed with the help of the hidden-gauge Lagrangians.
With the exchange potential obtained,   the $S$-wave bound states are searched for as the pole of the scattering amplitudes.

For the $D^{(*)}K^{(*)}$ system, all six isoscalar $S$-wave interactions produce  bound states while none of the isovector bound states are found in our calculation. In the bottom sector, analogous result can be found. We also make a calculation with the coupled-channel effect included. The results suggest a pattern generally consistent with the single-channel calculation. For the systems with a charm/antibottom and strange meson, the binding energies from the coupled-channel calculation is almost the same with these from the  single-channel calculation, and the state with $0^+$  near $D^*K^*$  threshold and state with $1^+$ near $DK^*$ threshold acquire widths.  For the systems with an  anticharm/bottom and strange meson, the cutoffs required to produce the states change a little larger than the systems with a charm/antibottom and strange meson, but still generally consistent with the single-channel calculation.

\begin{figure}[h!]
\centering
\includegraphics[scale=0.86,bb=50 100 350 250,clip]{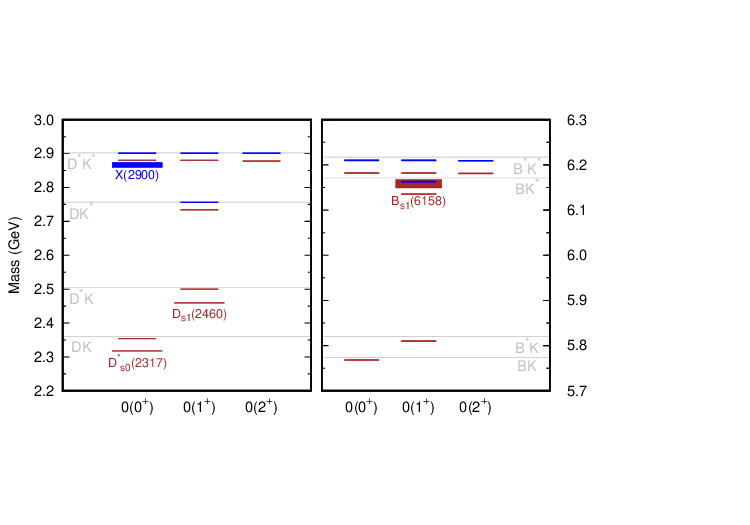}
\caption{The bound states from the interactions $D^{(*)}K^{(*)}/\bar{B}^{(*)}K^{(*)}$ (brown line) and $\bar{D}^{(*)}K^{(*)}/B^{(*)}K^{(*)}$ (blue line) at cutoff 1.4 and 2.0~GeV, respectively. The longer brown bars are for experimentally observed $D^*_{s0}(2317)$, $D_{s0}(2460)$, and $B_{sJ}(6185)$ with the height being the uncertainties of the experimental masses. The blue bar is for the $X(2900)$.}
\label{total}
\end{figure}

Since  the general pattern is not changed by coupled-channel effect and we only consider the couplings between the channels involved in the current work,  in Fig.~\ref{total} we still present the results with single-channel calculation for reference. Bound  states  produced from the $D^{(*)}K^{(*)}/\bar{B}^{(*)}K^{(*)}$ interaction are presented as the brown lines at a cutoff of 1.4~GeV.  The experimentally observed  charm-strange mesons $D^*_{s0}(2317)$ and $D_{s1}(2460)$ are also presented in the figure. Obviously, these two states can be related to the scalar and vector molecular states from the $DK$ and $D^*K$ states, respectively. With a cutoff of 1.4 GeV, the binding energies are smaller than the experimental values, which can be reproduced by choosing a larger cutoff, about 1.8~GeV.  With larger cutoff, the binding energies of other states should become larger also. In the coupled-channel calculation,  because the $D^*K$ threshold is the lowest one in the $D^*K-DK^*-D^*K^*$ system with $1^+$, no width is acquired by the state near the $D^*K$ threshold. It is analogous for the state near $DK$ threshold for $DK-D^*K^*$ system with $0^+$.  It is consistent with the small experimental widths of $D^*_{s0}(2317)$ and $D_{s1}(2460)$.  

\renewcommand\tabcolsep{0.04cm}
\renewcommand{\arraystretch}{1.2}
\begin{table}[h!]
\begin{center}
\caption{ The summary of the states obtained in the current work and relevant experimentally observed states. The $\Lambda$ is in units of GeV, and the theoretical (experimental) mass $M_{theo}$ ($M_{exp}$) and width $\Gamma_{theo}$ ($\Gamma_{exp}$) are in units of MeV. The experimental results are cited from~\cite{Zyla:2020zbs,Aaij:2020hon,Aaij:2020ypa,Aaij:2020hcw}.  Ratings are for reliability of the assignments. }
\label{Tab: Sum}
\begin{tabular}{ccccccccc}\toprule[2pt]\hline
System & $I(J^P)$ & $\Lambda$ & $M_{theo}$ & $\Gamma_{theo}$ & State& $M_{exp}$ & $\Gamma_{exp}$&Rating\\\hline
$D^{(*)}K^{(*)}$& $0(0^+)$ &1.7 &2313 & 0 & $D_{s0}^*(2317)$&$2317.8_{\pm0.5}$&$<3.8$ &***\\
                      & $$ & 1.7 &2859 & 28 & & & &   \\
                      & $0(1^+)$ & 1.7 &2466 & 0& $D_{s1}(2460)$&$2459.5_{\pm0.6}$&$<3.5$ &***\\
                      & $$ &1.7 &2656 &19 & & & &   \\
                      & $$ & 1.7 & 2839 &0& & & &   \\
                      & $0(2^+)$ &1.7 & 2834 &0& & & &   \\\hline
$\bar{D}^{(*)}K^{(*)}$& $0(0^+)$ & 2.7 & 2892 &82 & $X_0(2900)$ &$2866_{\pm7}$&$57.2_{\pm12.9}$& **\\
                              & $0(1^+)$ & 2.0 &2752 &7 & & & &  \\
                              & $$ & 2.0  &2901 &0& & & & \\
                              & $0(2^+)$ & 2.0 & 2901 &0 & & & &  \\\hline
$\bar{B}^{(*)}K^{(*)}$& $0(0^+)$ & 1.2 &5774 &0 & $$\\
                              & $$ & 1.2 & 6205& 2& $$ \\
                              & $0(1^+)$ &  1.2 & 5819 &0 & $$\\
                              & $$ &  1.2&  6158&2& $B_{sJ}(6158)$ &$6158_{\pm4\pm5}$ &$72_{\pm18\pm25}$&*\\
                              & $$ &1.2 &6194 &0 \\
                              & $0(2^+)$ & 1.2 & 6187 &0\\\hline
${B}^{(*)}K^{(*)}$& $0(0^+)$ & 1.8 & 6215 & 1& $$\\
                        & $0(1^+)$ & 1.8 &6169 &1& $$ \\
                        & $$ & 1.8 &6215&0 \\
                        & $0(2^+)$ & 1.8  &6214&0\\
\hline
\bottomrule[2pt]
\end{tabular}
\end{center}
\end{table}

The binding energies of the bottom bound states are generally larger than these in the charmed sector.  No partners of the $D^*_{s0}(2317)$ and $D_{s1}(2460)$ in the bottom sectors are observed experimentally. The $B_{s1}(5380)$ is near the $\bar{B}^*K$ threshold and carries a spin parity as an $S$-wave molecular state, but it is usually taken as a $1P$ $b\bar{s}$ state. Considering the closeness of $B_{s1}(5380)$ and the $\bar{B}^*K$ threshold, the mixing of  such state with the strong attractive  $\bar{B}^*K$ interaction should be taken seriously. For the two new $B_{sJ}$ states observed at LHCb recently,  the mass suggests it may correspond to a  state from the $\bar{B}K^*$ interaction if we assume it as a molecular state. The $B_{sJ}(6158)$  is closest to the $\bar{B}K^*$ threshold, and the decay channel $B^{*+}K^-$ leads to spin parity $1^+$ in $S$ wave. Hence, the $B_{sJ}(6158)$ is the best candidate of the $\bar{B}K^*$ molecular state with $0(1^+)$. However, two choices of two-peak structure were provided in the original experimental article; even a single resonance which can decay through both the $B^+K^-$ and $B^{*+}K^-$ channels is not excluded~\cite{Aaij:2020hcw}. If we consider that in both choices mass difference between the two peaks is found to be close to the $B^{*}-B$ mass difference of approximately 45 MeV, we even can not exclude possibility  that these two peaks are deeply bound states from interactions $\bar{B}^*K^*$ and $\bar{B}K^*$, respectively. Hence, more precise measurements of these states are required to make a determinative theoretical  analysis.

Since the  bound states from the systems $\bar{D}^{(*)}K^{(*)}$ and ${B}^{(*)}K^{(*)}$   with quark configuration $[\bar{Q}q][q\bar{s}]$ will not be mixed with the heavy-strange meson in the conventional quark model, it is easy to be distinguished and recognized as an exotic state.  As shown in Fig.~\ref{total}, our calculation suggests there are only four $\bar{D}^{(*)}K^{(*)}$ states produced from $S$-wave interaction, which are also all isoscalar. Three are from the $\bar{D}^{*}K^{*}$ interaction and one from the  $\bar{D}K^{*}$ interaction. After including the coupled-channel effect, a broad scalar state is produced near the $\bar{D}^*K^*$ threshold,  which is consistent with the experimental observed $X_0(2900)$. However, its binding energy seems a little smaller than the experimental value. More explicit analysis may be helpful to clarify such issue.

In Table~\ref{Tab: Sum}, we provide a summary of the results in the current work and compare with experimentally observed states. For the states which have experimental candidates,  our ratings of reliability of the assignments in the current model are also estimated as one to three stars based on the mass, width, cutoff, and the experimental status. For the $0(0^+)$ and $0(1^+)$ states from the $D^{(*)}K^{(*)}$ system, the masses and widths are close to the experimental values, and the cutoff is 1.7 GeV, which is reasonable in our model. The spin parities were determined in experiment. Hence, the assignments of $D_{s0}^*(2371)$ and $D_{s1}(2460)$ are considerably reliable.  For the $X_0(2900)$, the cutoff is obviously larger than other cases; hence, we rate such assignment as two stars. The spin parity of $B_{sJ}$ is not well determined in experiment, and the width cannot be well reproduced in the current model. The assignment needs more experimental information as input and theoretical studies. Other predicted state are also summarized and the mass and width with  chosen cutoff are also provided for reference.

In summary, the $D^{(*)}K^{(*)}/\bar{D}^{(*)}K^{(*)}$ and $\bar{B}^{(*)}K^{(*)}/B^{(*)}K^{(*)}$ systems are studied in a qBSE approach.  Ten isoscalar molecular states are produced from $S$-wave interaction in the  charmed and bottom sectors, respectively.  Further experimental information is very helpful to understand such spectrum of the molecular state, and the origin of the $X(2900)$ and new $B_{sJ}$ states.

\vskip 10pt

\noindent {\bf Acknowledgement} This project is supported by the National Natural Science
Foundation of China (Grant No. 11675228).

\end{document}